\documentclass{aa}
\usepackage{graphicx}
\begin{document}
%

   \headnote{Research Note}
   \title{First NACO observations of the Brown Dwarf LHS 2397aB\thanks{Based on observations obtained at ESO-Paranal (NACO) and HST.}}


   \author{E. Masciadri \inst{1}, W. Brandner \inst{1}, H. Bouy \inst{2}, R. Lenzen \inst{1},
A.M. Lagrange \inst{3} \and F. Lacombe \inst{4}}

   \offprints{Masciadri E. - \email{masciadri@mpia.de}}

   \institute{Max-Plank Institut f\"ur Astronomie, K\"onigstuhl 17, 
              D-69117 Heidelberg, Germany \\
              \and
              European Southern Observatory, D-85748 Garching bei M\"unchen, Germany \\
              \and Laboratoire d'Astrophysique, Observatoire de Grenoble, 414,  
              Rue de la Piscine, BP 53, 38041 Grenoble \\ Cedex 9, France \\
              \and 
              Observatoire Paris-Meudon, LESIA, Place Jules Janssen,
              92195 Meudon Cedex, France}

   \date{Received 27 March 2003/ Accepted 19 August 2003}

   \abstract{Observations of the standard late type M8 star LHS 2397aA were obtained at the ESO-VLT 8m telescope ``Yepun'' using the NAOS/CONICA Adaptive Optics facility. The observations were taken during the NACO commissioning, and the infrared standard star LHS 2397aA was observed in the H, and Ks broad band filters. In both bands the brown dwarf companion LHS2397aB was detected. Using a program recently developed (Bouy et al., 2003) for the detection of stellar binaries we calculated the principal astrometric parameters (angular binary separation and position angle P.A.) and the photometry of LHS 2397aA and LHS 2397aB. Our study largely confirms previous results obtained with the AO-Hokupa'a facility at Gemini-North (Freed et al., 2003); however a few discrepancies are observed.
          }
\authorrunning{Masciadri}   
\maketitle
\markboth{First NACO observations of the Brown Dwarf LHS 2397aB}{Masciadri et al.}
%

\section{Introduction}

The study of the frequency of brown dwarf (BD) companions at different distances from the primary is a crucial point for the determination of the origin of these sub-stellar objects ($\sim$ [$13$ - $75$] M$_{J}$). The most popular scenarios are: (1) the BDs are the result of a cloud fragmentation mechanism (Elmegreen, 1999). After the formation of a nucleus, the BDs would be built by accretion from a disc like star, (2) the BDs are the result of a disk fragmentation mechanism (Pickett et al., 2000), (3) the BDs are ejected stellar embryos (Reipurth $\&$ Clarke, 2001). In the last years several BDs companions of Sun-like stars (Marcy et al. (2003), Gizis et al., 2001, Kirkpatrick et al. (2001), Potter et al. (2002)) and of very low mass (VLM) stars and/or other BDs (M\'artin et al. (2000), Koerner et al. (1999), Bouy et al. (2003), Close et al. (2003), Burgasser et al. (2003)) were discovered. These discoveries can provide useful information for a statistical analysis of the BD binary characteristics (orbital parameters, mass ratio, and so on) and BD binary frequency as a function of the separation. 
We refer the reader to the papers by Bouy et al. (2003), Close et al. (2003) and Burgasser et al. (2003) for more complete discussions of the subject using the most recent statistical results.\newline\newline
LHS 2397aA ($11$:$21$:$49$, $-13$:$13$:$08$) is a very low mass star 
(spectral type M8) at a distance 
of $14.3$ $pc$. This allows us to study it at resolutions
of the order of a few AUs ($10^{-2}$ arcsec). Optical and NIR 
photometry was reported by Leggett et al. (2002) and the rotation velocity 
($v\sin{i}$ $=$ $20$ $km s^{-1}$) and the 
chromospheric (H$_{\alpha}$ - EW $=$ $29.4$ \AA) activity was studied 
by Mohanty $\&$ Basri (2003). 
A BD companion to LHS 2397aB was discovered by Freed et al. (2003) 
at a close distance of only $\sim$2.86 AU.  
In this paper we report the astrometric parameters and the photometry as 
retrieved by observations done with the ESO-VLT 8m telescope ``Yepun'' 
and NACO, which consists of the facility adaptive optics system NAOS 
(Rousset et al.\ 2002) and the infrared camera CONICA (Lenzen et al.\ 1998).
Our results will be discussed taking into account the previous observations 
of the same object: images obtained with the adaptive optics system Hokupa'a, 
at Gemini North (Freed et al., 2003) and images from the HST/WFPC2 archive 
data (G06345, P.I. Kirkpatrick).

\begin{table*}
\begin{center}
\begin{tabular}{lccc}
\hline
\noalign{\smallskip}
\hline
\noalign{\smallskip}
  &  & This paper & Freed et al. \\
\hline
HST/WFPC2 1997/4/12   & Separation ($\arcsec$) & 0.255 $\pm$  0.007 &  0.270 $\pm$  0.010 \\
 &  Position Angle (degrees) & 90.3 $\pm$ 0.8 & 82.8 $\pm$ 1.5    \\
 &  $\Delta{Ic}$ & 4.34 $\pm$ 0.14 & 4.42 $\pm$  0.17   \\
\hline
\noalign{\smallskip}
AO-HOKUPA 2002/2/7 &  Separation ($\arcsec$) & - & 0.207 $\pm$  0.007 \\

 & Position Angle (degrees) & - & 151.9 $\pm$ 1.2 \\
 &  $\Delta{H}$ & - &   3.15 $\pm$ 0.30    \\
 &  $\Delta{Ks}$ & - &  2.77 $\pm$ 0.10     \\
\hline
\noalign{\smallskip}
 AO-NACO  2002/3/26 & Separation ($\arcsec$) & 0.164 $\pm$  0.004 & -  \\
 & Position Angle (degrees) & 157.3 $\pm$ $2.0$ & -\\
 &  $\Delta{H}$ & 3.03 $\pm$ 0.13&   -    \\
 &  $\Delta{Ks}$ & 2.53 $\pm$ 0.03 &  -  \\
\hline
\noalign{\smallskip}
\hline
\noalign{\smallskip} 
\end{tabular}
\caption{Summary of the astrometric and photometric estimations. 
In the first three lines the HST/WFPC2 results
from the Masciadri et al. (2003) and Freed et al. (2003) analyses are shown. 
In the following lines the results obtained from AO-Hokupa'a data and 
the from AO-NACO data are shown.}
\label{table_period}
\end{center}
\end{table*} 

\begin{figure*}[H]
\centering
\includegraphics[width=15cm]{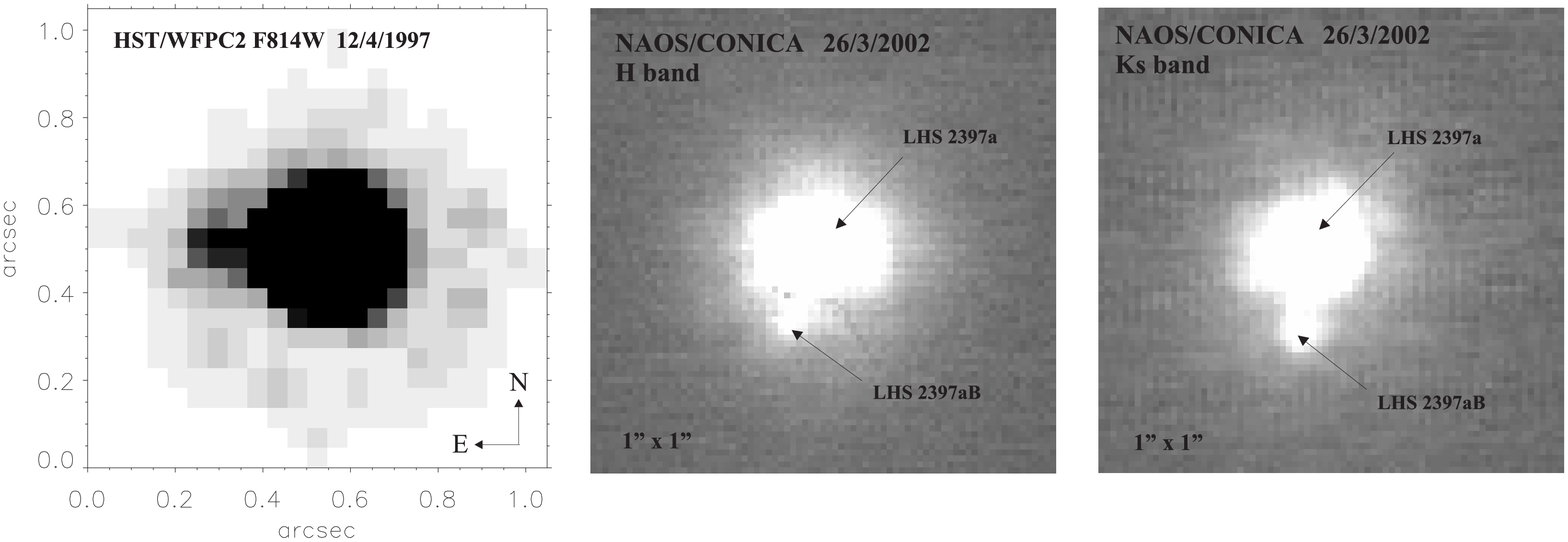}
\caption{Left: LHS 2397aA and LHS 2397aB images taken 
with the HST/WFPC2 camera and the F814W filter 
on $12/4/1997$. Centre and right: LHS 2397aA and LHS 2397aB images taken in H (center) 
and Ks (right) broad band with AO-NACO on $26/3/2002$. The field of view (FOV) of all 
the images is $1''$ x $1''$.}
\label{fig1} 
\end{figure*}

\section{Observations}

Observations of LHS 2397aA and other infrared standard stars were obtained on 
March 26, 2002.  The observations were obtained during the 
NACO commissioning, and the infrared standard 
star LHS 2397aA was observed in the H and Ks broad band filters with 
the CONICA S13 camera. LHS 2397aA was observed in a
5-point dither pattern with individual exposure times of 5 $\times$ 2s 
(i.e.\ 10s per dither position) for the S13 camera (pixel equivalent to 
$0\farcs{013}$ and $14''$ x $14''$ FOV). Observations of other
infrared standard stars (FS 18 and FS 20) were obtained in a similar
fashion i.e. with the same exposure time and dither position.

A precise knowledge of the pixel scale and and on-sky orientation 
(instrumental position angle) of CONICA
is a pre-requisite for a sound determination of the separation and 
position angle of the LHS2397aA binary.
The lab and on-sky calibration of these parameters is summarized in Brandner (2002).
During the commissioning, two independent data sets were obtained to get 
precise on-sky
pixel scales and position angles for CONICA. First, astrometric binaries were observed. 
For each binary, instrumental positions (pixel coordinates) were compared with true 
separation and position angle. Secondly, the Galactic Center region was observed 
multiple times during the 
commissioning with NACO. Here a global astrometric solution was obtained based on 
the positions of maser sources in the Galactic
Center region (Sch\"odel et al. (2002), Reid et al., (2003)). 
The calibrations obtained with both methods were quite consistent, and yielded for 
camera S13 the on-sky pixel scale with a precision of $0.22$$\%$ 
($13.26$ $\pm$ $0.03$ $mas$) and the instrumental position angle with a precision of 
$\pm$ $0.5$ $deg$.

Our observations were done with a seeing in the range $0\farcs{68}$ - $0\farcs{74}$, 
an airmass of $\sim$ $1.15$ and resulting FWHM equal to 
$\sim$ $0\farcs{078}$ in the K band. We selected the PSF stars which were observed under similar conditions 
(equivalent seeing, comparable FWHM). 
Figure \ref{fig1} shows images of LHS 2397aA 
taken in the H (centre) and Ks (right) broadbands over $1''$ x $1''$ FOV.
The grey table is logarithmic, no low-frequency filtering was done (as in the equivalent 
images in Freed et al., 2003 - Fig.1). Figure \ref{fig1} (left) shows the image taken with 
the HST/WFPC2 camera in F814W that we reduced to complete our analysis. 

\section{Data analysis}

A PSF fitting program, conceived to compute the separation, the position angle 
and the flux ratio of stellar binary systems was used to reduce our data. 
This program (Bouy et al., 2003) is based on a fit of observed PSF stars 
trying to reproduce the observed binary system. 
A detailed description as well as a validation of the technique 
is described in Bouy et al. (2003).
The fitting program is applied to square fields centred on both the scientific 
targets and the standard stars. 
The square field measures $64$ x $64$ pixels (corresponding to 
$0\farcs{8}$ x $0\farcs{8}$).  
We note that the FS 18 standard is a binary with a separation 
of $1\farcs{36}$ i.e. 
a distance larger than the 
size of the square field on which the fitting program is applied. 
This means that 
the binarity of FS 18 does 
not present a problem for our data reduction. 
We first calculated, in an iterative way, the optimized model 
(i.e. a synthetic binary) and then we 
retrieved a rms ($\sigma_{res}$) of the residuals
as a quantitative measure on the quality of our fit. 
The calculation of the astrometric parameters (separation and P.A.) 
was done for the 
Ks band data in which the binary was well resolved and the SNR
was good enough to retrieve reliable values for the astrometric parameters.
For the H band data the astrometric parameters derived from the Ks band data 
were fixed and only the brightness ratio was left as a free parameter.

Table \ref{table_period} shows the astrometric and photometric values 
retrieved from our AO-NACO data taken in March $2002$, by 
Freed et al. (2003) (February $2002$ - AO-Hokupa'a) and by HST/WFPC2 (April $1997$).  
Both the astrometric and photometric NACO 
values shown in Table \ref{table_period} 
are obtained considering the systematic (see further down on the calibration 
procedure) and statistical errors.
The angular separation of $0\farcs{164}$ $\pm$ $0\farcs{004}$ translates
to a physical separation of $2.34$ $\pm$ $0.14$ AU.. 
The absolute magnitude was calculated assuming 
a distance of $d$ $=$ $14.3$ $\pm$ $0.4$ $pc$.

The procedure used to calculate the fitting parameters is as follows.
For each observation of LHS 2397aA, two sets of "corresponding" PSFs stars
(observations of FS 18 and FS 20 with similar SR and FWHM as LHS 2397aA)
were used to model the observed brightness distribution,
and hence to derive the parameters for separation, position angle and flux
ratio. The final numbers reported in this paper were computed based on a
weighted average of each fitting parameter: mean separation
($\overline{sep.}$), position angle ($\overline{pa}$) and flux ratio
($\overline{q}$). The weights of the statistical estimations
are inversely proportional to the rms of the residuals ($\sigma_{res_i}$) 
model-observations 
found in the fitting process. The index $i$ refers to the library of 
images and different PSFs.

To calibrate the program we applied it to artificial binaries covering 
angular separations
including the range [$0\farcs{1}$-$0\farcs{4}$] and  
differences in magnitude between the components 
in the [$0.2$-$3.2$] $mag$ range. We note that the $\Delta{m_{Ks}}$ retrieved 
by our fit before the calibration is equal to $2.17$ $mag$ so it belongs 
to the studied range. A detailed description of the calibration method 
can be found in Section 2.3.2 in Bouy et al. (2003). 
The calibration procedure permitted us to find the off-sets i.e. 
a sort of systematic error for each parameter (astrometric and 
photometric) that we have to add to our statistic estimations.

The results of the calibration procedure gives us an off-set for the 
angular separation equal to $3.2$ $\pm$ $0.003$ mas, 
the systematic error for the P.A. 
is estimated equal to $3.4$ $\pm$ $2$ degrees 
and that for the magnitude difference 
is $0.36$ $\pm$ $0.02$ $mag$ ($\Delta{m_{Ks}}$) and 
$0.69$ $\pm$ $0.06$ $mag$ ($\Delta{m_H}$).
We note that the calibration results are instrument dependent and our 
estimations are obtained with a not too rich sample (a few PSFs). 
It is planned to improve the calibration for NACO (an observing 
program whose task is the search for BDs binaries 
is in progress with NACO) using a richer sample of PSFs. 
At the present time we can state that a detailed calibration was been 
done for HST observations (Bouy et al., 2003) 
using a rich statistical sample. 

Due to the quite high SR 
of the HST/WFPC2  images, the binary is well resolved and 
systematic effects in the fitting procedure are small. This means that
in this case we do not need a calibration for the program 
(see Bouy et al. 2003).
We calculated averaging the values of the four best fits (i.e. the smallest residuals) 
out of nine fits obtained using as the PSF the following stars: BRI 0021, LHS 2243, 
TVLM 868 and RG 0050.

Table \ref{table_phot2} shows, for the primary (LHS 2397aA) and the 
companion (LHS 2397aB), the apparent and absolute magnitude 
in the H and Ks bands (NACO and Hokupa'a data) and $I_{c}$ band (HST/WFPC2 data). The HST handbook states that the F814W filter is a very close approximation to the Johnson-Cousins I$_{c}$ and colour terms between these filters are very small. We can hence assume that F814W magnitudes are a good approximation of the I$_{c}$ magnitudes. We consider (see Table\ref{table_period}) $\Delta(I_{c})$ $=$ $\Delta{F814W}$ + $0.06$ $mag$ as in the Freed paper. 

\section{Discussion}
\label{disc}

\begin{table}
\begin{center}
\begin{tabular}{lcc}
\hline
\noalign{\smallskip}
\hline
\noalign{\smallskip}
 & $LHS 2397aA$& $LHS 2397aB$ \\
\hline
\noalign{\smallskip}
$m_{Ic}$ & 14.81 $\pm$ 0.01 & 19.08 $\pm$ 0.14 \\
$m_{Ic}^{*}$ & 15.07 $\pm$ 0.03 & 19.49  $\pm$ 0.17 \\
$m_{H}$ & 11.32 $\pm$ 0.02 & 14.35 $\pm$ 0.25 \\
$m_{H}^{*}$ & 11.32 $\pm$ 0.05 & 14.47  $\pm$ 0.30 \\
$m_{Ks}$ &10.82 $\pm$ 0.02 &  13.36 $\pm$ 0.02   \\
$m_{Ks}^{*}$ & 10.80 $\pm$ 0.03  &  13.57 $\pm$ 0.10  \\
\hline
$M_{Ic}$ & 14.03 $\pm$ 0.06 & 18.30  $\pm$ 0.15 \\
$M_{Ic}^{*}$ & 14.29   $\pm$ 0.07 & 18.71 $\pm$ 0.18 \\
$M_{H}$ & 10.56$\pm$ 0.14 &  13.58  $\pm$ 0.18 \\
$M_{H}^{*}$ & 10.54  $\pm$ 0.08 &13.69 $\pm$0.31 \\
$M_{Ks}$ & 10.05 $\pm$ 0.14 &  12.58 $\pm$ 0.14\\
$M_{Ks}^{*}$ & 10.03 $\pm$ 0.07 & 12.80  $\pm$ 0.12 \\
\hline
\noalign{\smallskip}
\hline
\noalign{\smallskip} 
\end{tabular}
\caption{Apparent (first 6 lines) and absolute (last 6 lines) magnitude calculated in the H, Ks and Ic band. The values marked with an asterisk refer to the Freed et al. (2003) paper.}
\label{table_phot2}
\end{center}
\end{table}

The photometric results retrieved in our analysis are in good agreement with Freed et al. (2003). 
The few discrepancies are not larger than the error bar with 
which the $\Delta(m)$ is estimated for the ground-based observations. 
Our estimation of $\Delta(m_{Ic})$ is equal to $4.34$ $\pm$ $0.14$ $mag$ in good agreement with that estimated by Freed et al. ($4.42$ $\pm$ $0.17$ $mag$).\newline

The parameter {\it ORIENTAT} in the fits file header of the HST/WFPC2 image gives the position of the north with respect to the y axis. Using this offset angle, and the measured instrumental $P.A._{inst}$ $=$ $320$$^{\circ}$.$3$, we get $P.A.$ $=$ {\it ORIENTAT} - ($360$$^{\circ}$ - $P.A._{inst}$) $=$ $90^{\circ}$$.3$. 
We note that this differs from the $82^{\circ}$$.8$ reported by Freed et al. (2003). 
Since the binary is well resolved, the discrepancy is most likely due to a problem in the transformation from instrumental to true P.A. as done by Freed et al. 

We note that the angular separation measured in the different periods seems to decrease with the time. This could indicate an inclined elliptical orbit of the companion. Nevertheless, the epoch difference between our observations (AO-NACO) and the one by Freed et al. (AO-Hokupa'a) is just one month. This seems to be a quite small temporal difference to justify a difference of about $0.''04$ and we suggest further observations to confirm the trend or to get a better estimate of the angular separation. 
Besides this we note that, because of the systematic error, the uncertainty on the position angle is about $2$ degrees. We think that it is premature to try to retrieve an orbit trajectory. Under the hypothesis of a circular orbit the companion should complete a revolution in about $26.5$ yr ($13.6$ deg/yr). More reliable orbital parameters will become available with further observations to be made in the incomin years.

\section{Conclusions}

The most important conclusion of our study is that the analysis of observations of LHS 2397aA taken in March 2002 with AO-NACO gives astrometric and photometric results in good agreement with the previous ones obtained with another ground-based AO-system (Hokupa'a) although the reduction of data was made 
in two different ways. The difference in the photometry is smaller than the uncertainty of the measurements. In conclusion we can confirm the results of Freed et al. (2003) for photometry, age and spectral type. We suggest further observations to confirm the angular separation of $0\farcs{164}$ estimated in our analysis and to attemp a calculation of the companion orbit. We note that the P.A. derived by us for the HST data differs from the P.A. quoted by Freed et al. Our determination suggests a longer orbital period.

\begin{acknowledgements}
These data are part of the NACO commissioning data. We acknowledge the staff who took part in the observations. We acknowledge the anonymous referee for the useful commentaries and Dave Butler for the improvements suggested in the English form. 

\end{acknowledgements}


\begin{thebibliography}{}

\bibitem[2003]{Bouy2003}
Bouy H. et al., 2003,
  {\it AJ}, accepted

\bibitem[2003]{Brandner2003}
Brandner W., 2000, 
   {\it NACO Commissioning Report - Comm 4 VLT-TRE-ESO-14200-2817}

\bibitem[2003]{Burgasser2003}
Burgasser A.J., Kirkpatrick J.D., Reid I.N., Brown M.E., Miskey C.L., 2003,
  {\it ApJ}, {\bf 568}, 512
  
\bibitem[2003]{Close2003}
Close et al., 2003,
  {\it ApJ}, to be published on April, 2003


\bibitem[1999]{Elmegreen1999}
Elmegreen B. G., 1999,
  {\it ApJ}, {\bf 522}, 915

\bibitem[2003]{Freed2003}
Freed M., Close L.M., Siegler N., 2003,
  {\it ApJ}, {\bf 584}, 453

\bibitem[2001]{Gizis2001}
Gizis J., Kirkpatrick J.D., Burgasser A., Reid I.N., Monet D.G., Liebert J. $\&$ Wilson J.C., 2001,
  {\it ApJ}, {\bf 551},L163

\bibitem[2002]{Legget2002}
Legget S.K. et al., 2002, 
  {\it ApJ}, {\bf 564}, 452

\bibitem[2002]{Lenzen2002}
Lenzen R. et al., 1998, 
  {\it SPIE}, {\bf 3354}, 606

\bibitem[2001]{Kirkpatrick12001}
Kirkpatrick J.D., Dahan C.C., Monet D.G., Reid I.N., Gizis J.E., Liebert J., Burgasser A.G., 2001, 
  {\it AJ}, 121, 3235

\bibitem[1999]{Koerner1999}
Koerner D.W., Kirkpatrick J.D., McElwain M.W., Bonaventura N.R., 1999, 
  {\it ApJ}, 526, L25

\bibitem[2003]{Marcy2003}
Marcy G., Butler R.P., Fischer D.A. $\&$ Vogt S.S., 2003, 
  {\it ASP Conf. Ser. Scientific Frontiers in Research on Extrasolar Planets}, Ed. Deming D. $\&$ Seager S., in press

\bibitem[2003]{Mohanty2003}
Mohanty S., Basri G., 2003, 
{\it ApJ}, {\bf 583}, 451

\bibitem[2000]{Pickett2000}
Pickett B.K., Durisen R.H., Cassen P., Mejia A.C.,  2000, 
{\it ApJ}, {\bf 540}, L95

\bibitem[2002]{Potter2002}
Potter D., Martin E.L., Cushing M.C., Baudoz P., Brandner W., Guyon O., Neuh\"auser R., 2002, 
 {\it ApJ}, {\bf 567}, 133

\bibitem[2003]{Reid2003}
Reid et al., 2003
  {\it ApJ}, {\bf 587}, 208

\bibitem[2001]{Reipurth2001}
Reipurth$\&$ Clarke 2001, 
{\it AJ}, {\bf 122}, 432

\bibitem[2000]{Rousset2000}
Rousset G. et al., 2000, 
{\it SPIE}, {\bf 4007}, 72

\end{thebibliography}
\end{document}